\documentclass[fleqn,twoside]{article}
\usepackage{amsmath,amssymb,espcrc2,graphicx}

\newcommand{\dif}{\mathrm{d}}
\newcommand{\xB}{x_{\scriptscriptstyle{B}}}
\newcommand{\Pom}{{\hspace{ -0.1em}I\hspace{-0.2em}P}}
\newcommand{\Reg}{{\hspace{ -0.1em}I\hspace{-0.2em}R}}
\newcommand{\xPom}{x_\Pom}
\newcommand{\chisq}{\chi^2/\mathrm{d.o.f.}}

\title{Simultaneous QCD analysis of diffractive and inclusive DIS data}

\author{\underline{G. Watt}\address[IPPP]{Institute for Particle Physics Phenomenology, University of Durham, DH1 3LE, UK},
  A.D. Martin\addressmark[IPPP]
  and
  M.G. Ryskin\addressmark[IPPP]$^,$\address{Petersburg Nuclear Physics Institute, Gatchina, St.~Petersburg, 188300, Russia}}

\begin{document}

\begin{abstract}
We present a novel analysis of diffractive DIS data, in which the input parton distributions of the Pomeron are parameterised using perturbative QCD expressions.  In addition to the usual two-gluon model for the perturbative Pomeron, we allow for the possibility that it may be made from two sea quarks.  In particular, we treat individually the components of the Pomeron of different size.  This property allows the absorptive corrections to the inclusive DIS structure function $F_2$ to be calculated using the AGK cutting rules.  The absorptive effects are found to enhance the size of the gluon distribution of the proton at small $x$.
\vspace{1pc}
\end{abstract}

\maketitle

\section{DIFFRACTIVE DIS}

A notable feature of deep-inelastic scattering is the existence of diffractive events, $\gamma^* p\to X p$, in which the slightly deflected proton and the cluster $X$ of outgoing hadrons are well-separated in rapidity.  The large rapidity gap is believed to be associated with Pomeron exchange.  The diffractive events make up an appreciable fraction ($\approx10$\%) of all (inclusive) deep-inelastic events, $\gamma^* p \to X$.  We will refer to the diffractive and inclusive processes as DDIS and DIS respectively.

The diffractive structure function, $F_2^{D(3)}(\xPom,\beta,Q^2)$, has recently been measured to high precision by the ZEUS \cite{Capua} and H1 \cite{Schaetzel} Collaborations at HERA \cite{Bruni}.

Both HERA experiments have analysed their DDIS data using the `resolved Pomeron' model \cite{Ingelman:1984ns}, where the $\xPom$ dependence of $F_2^{D(3)}$ is assumed to factorise into a `Pomeron flux factor' and the Pomeron is treated as having parton distribution functions (PDFs) just like a hadron.  Since the Pomeron is not a particle, this model must be taken on a purely phenomenological basis.

An alternative description of DDIS is provided by interpreting Pomeron exchange as two-gluon exchange, such as in the BEKW model \cite{Bartels:1998ea}.  However, the BEKW parameterisation is rather far from the original perturbative QCD (pQCD) calculations of W\"usthoff \cite{Wusthoff:1997fz}.

\section{NEW pQCD APPROACH TO DDIS}
\begin{figure}[!tb]
  \begin{center}
    \begin{minipage}{0.23\textwidth}
      \hspace{0.35\textwidth}(a)\\
      \includegraphics[width=\textwidth]{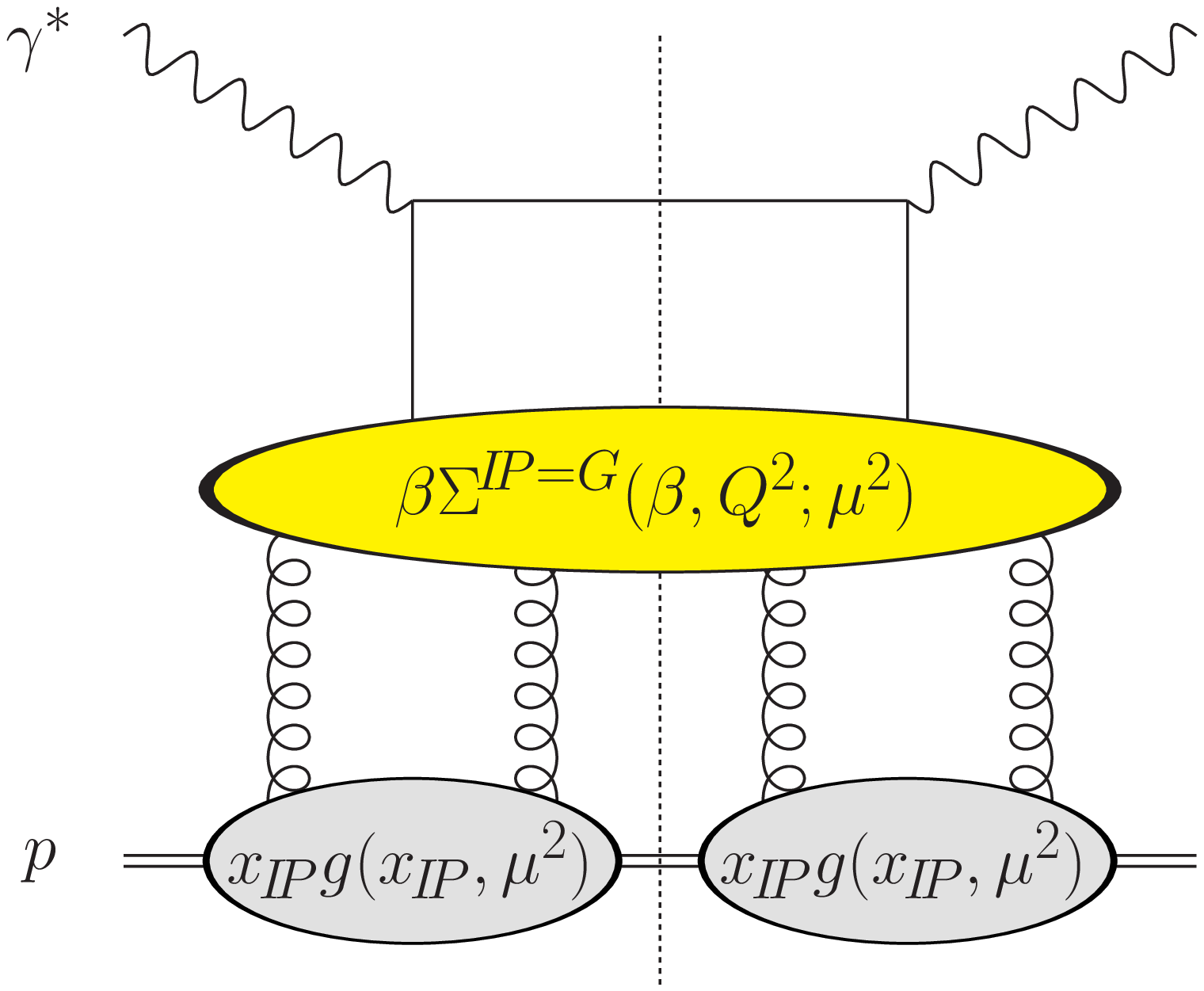}
    \end{minipage}%
    \hfill
    \begin{minipage}{0.23\textwidth}
      \hspace{0.35\textwidth}(b)\\
      \includegraphics[width=\textwidth]{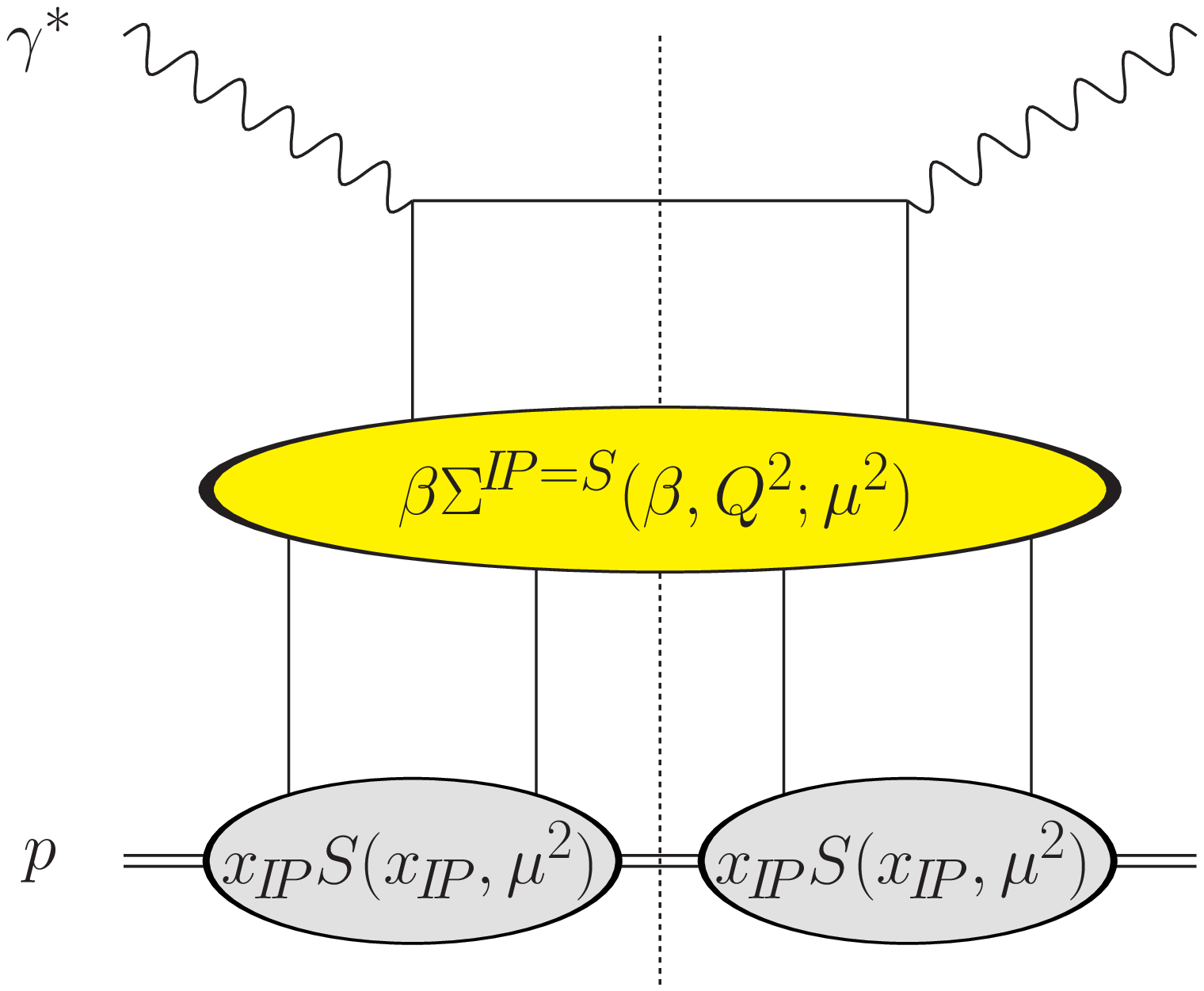}
    \end{minipage}
  \end{center}
  \caption{New pQCD approach to DDIS.  Each component of the perturbative Pomeron of size $1/\mu$ is represented by either (a) two $t$-channel gluons in a colour singlet or (b) sea quark--antiquark exchange.  The perturbative Pomeron flux factors $f_{\Pom}(\xPom;\mu^2)$ are given in terms of the gluon and sea quark distributions of the proton, $g(\xPom,\mu^2)$ and $S(\xPom,\mu^2)$.  The Pomeron structure function $F_{2}^{\Pom}(\beta,Q^2;\mu^2)$ is evaluated at NLO from the quark singlet, $\Sigma^\Pom(\beta,Q^2;\mu^2)$, and gluon, $g^\Pom(\beta,Q^2;\mu^2)$, distributions of the Pomeron.}
  \label{fig:f2d3pom}
\end{figure}

In pQCD, it is known that the Pomeron singularity is not an isolated pole, but a branch cut, in the complex angular momentum plane \cite{Lipatov:1985uk}.  The pole singularity corresponds to a single particle, whereas a branch cut may be regarded as a continuum series of poles.  That is, the Pomeron wave function consists of a continuous number of components.  Each component $i$ has its own size, $1/\mu_i$.  The DGLAP evolution of a component should start from its own scale $\mu_i$, provided that $\mu_i$ is large enough for the perturbative evolution to be valid.  Therefore, the expression for the diffractive structure function $F_2^{D(3)}$ contains an integral over the Pomeron size, or rather over the scale $\mu$:
\begin{multline}
  \label{eq:F2D3P}
  F_{2,{\rm P}}^{D(3)}(\xPom,\beta,Q^2) = \sum_{\Pom=G,S,GS}\;\int_{Q_0^2}^{Q^2}\!\dif\mu^2\;\\\times f_{\Pom}(\xPom;\mu^2)\; F_2^\Pom(\beta,Q^2;\mu^2).
\end{multline}
Here, the subscript ${\rm P}$ on $F_{2,{\rm P}}^{D(3)}$ is to indicate that this is the perturbative contribution with $\mu>Q_0\sim 1$ GeV.  The notation $\Pom=G,S,GS$ denotes that the perturbative Pomeron is represented by two $t$-channel gluons, two $t$-channel sea quarks, or the interference between these, respectively; see Fig.~\ref{fig:f2d3pom}.  The necessity for the `two-quark Pomeron' in addition to the two-gluon Pomeron is due to the valence-like shape of the gluon distribution of the proton at low scales.  The $\beta$ dependence of the input Pomeron PDFs, $\Sigma^\Pom(\beta,\mu^2;\mu^2)$ and $g^\Pom(\beta,\mu^2;\mu^2)$, is obtained from the lowest-order Feynman diagrams, with the normalisations allowed to go free in fits to the DDIS data to account for higher-order QCD corrections.

In addition to the perturbative contribution \eqref{eq:F2D3P} to $F_2^{D(3)}$, we also include a non-perturbative (${\rm NP}$) contribution, a twist-four contribution, and a secondary Reggeon ($\Reg$) contribution.  For more details, see \cite{Martin:2004xw}.

\section{ABSORPTIVE CORRECTIONS TO $F_2$}

\begin{figure}[!tb]
  \centering
  \includegraphics[width=0.5\textwidth]{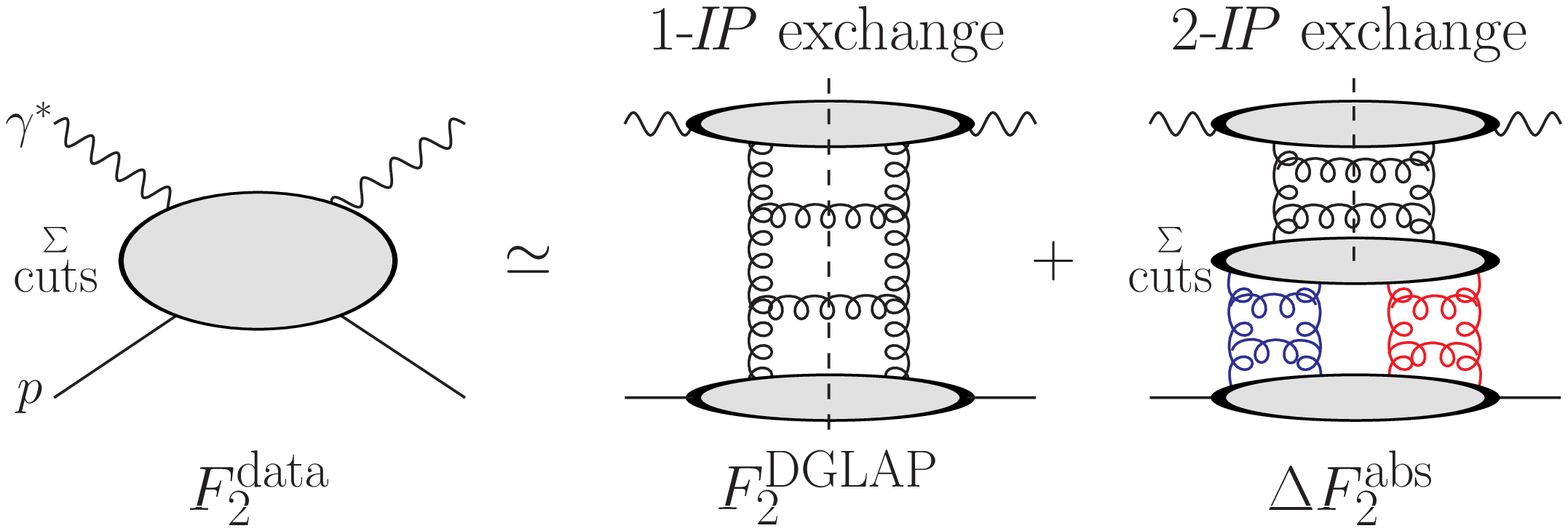}\\[5mm]
  \includegraphics[width=0.5\textwidth]{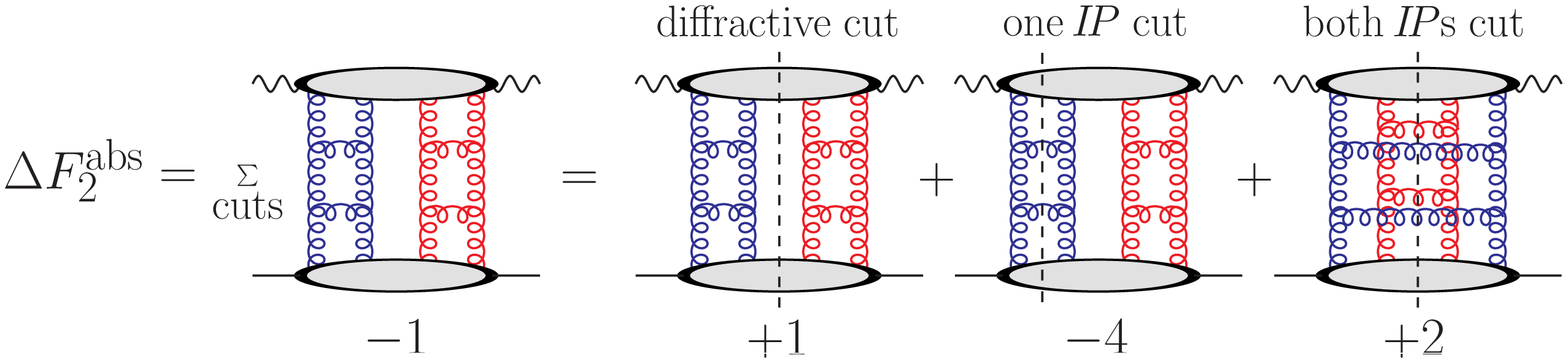}
  \caption{Absorptive corrections to $F_2$ due to the $\Pom\otimes\Pom$ contribution, and application of the AGK cutting rules \cite{Abramovsky:1973fm}.}
  \label{fig:absorpcorr}
\end{figure}
The total proton structure function, $F_2^{\mathrm{data}}(\xB,Q^2)$, as measured by experiment, can be approximately written as a sum of the one-Pomeron (DGLAP) contribution and absorptive corrections due to two-Pomeron exchange; see Fig.~\ref{fig:absorpcorr}.  Application of the AGK cutting rules \cite{Abramovsky:1973fm} to the $\Pom\otimes\Pom$ contribution gives
\begin{equation*}
  \Delta F_2^\mathrm{abs}(\xB,Q^2) = - \int_{Q_0^2}^{Q^2}\!\dif{\mu^2}\;F_2^D(\xB,Q^2;\mu^2),
\end{equation*}
where $F_2^D(\xB,Q^2;\mu^2)$ is the contribution to the diffractive structure function $F_2^{D(3)}$ (integrated over $\xPom$) which originates from a perturbative component of the Pomeron of size $1/\mu$.  We can thus separate the screening corrections coming from low $\mu<Q_0$, which are included in the input parameterisations taken at a scale $Q_0$, from the absorptive effects at small distances ($\mu>Q_0$) which occur during the DGLAP evolution in the analysis of DIS data.  For more details, see \cite{Martin:2004xx}.

\section{SIMULTANEOUS $F_2$ + $F_2^{D(3)}$ FIT}

The `simultaneous' fit of DIS and DDIS data proceeds as follows:
\begin{enumerate}
  \renewcommand{\labelenumi}{(\roman{enumi})}
\item Start by fitting ZEUS \cite{ZEUSF2} and H1 \cite{H1F2} $F_2$ data (279 points) with no absorptive corrections, similar to the MRST2001 NLO analysis \cite{MRST2001}.
\item Fit ZEUS \cite{ZEUSdata} and H1 \cite{H1data} $F_2^{D(3)}$ data (404 points) using $g(\xPom,\mu^2)$ and $S(\xPom,\mu^2)$ from the previous $F_2$ fit.
\item Fit $F_2^{\mathrm{DGLAP}} = F_2^{\mathrm{data}} + \left\lvert\Delta F_2^{\mathrm{abs}}\right\rvert$, with $\Delta F_2^{\mathrm{abs}}$ from the previous $F_2^{D(3)}$ fit.
\item Go to (ii).
\end{enumerate}
In practice, only a few iterations of steps (ii) and (iii) are needed for convergence.

\begin{figure}[!tb]
  \centering
  \includegraphics[width=0.5\textwidth]{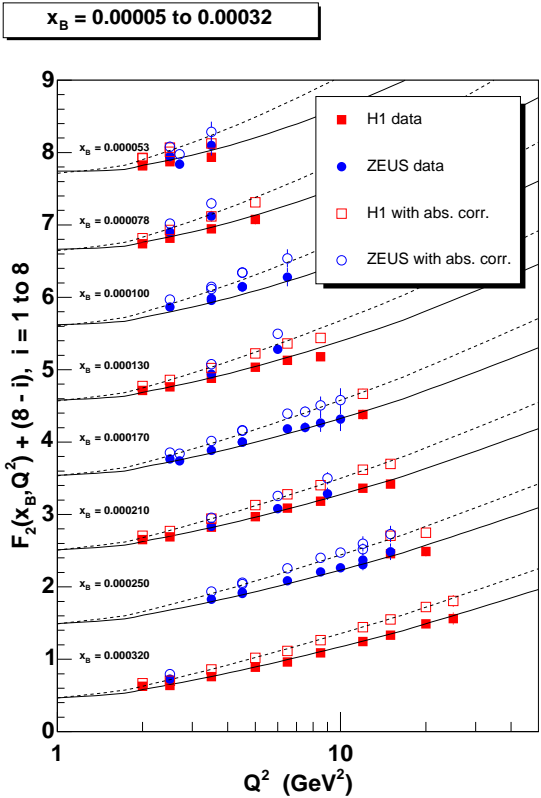}
  \caption{$F_2$ data \cite{ZEUSF2,H1F2} at small $\xB$ before and after absorptive corrections have been included.  Only data points included in the fits, shown by the solid and dashed lines, are plotted.}
  \label{fig:f2data}
\end{figure}

In Fig.~\ref{fig:f2data} we show the $F_2$ data at the smallest $\xB$ values, before and after the absorptive corrections have been applied.  The predictions of the corresponding fits, shown by the solid and dashed curves, are also plotted.
\begin{figure}[!tb]
  \centering
  \includegraphics[width=0.5\textwidth,clip]{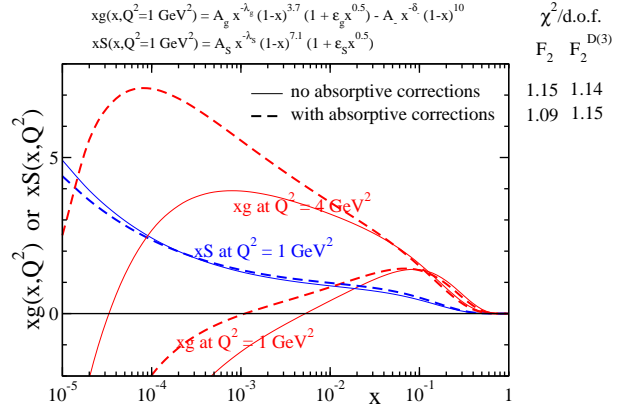}
  \caption{The gluon and sea quark distributions obtained from a NLO DGLAP fit to $F_2$, before and after absorptive corrections have been included.}
  \label{fig:neggluon}
\end{figure}
In Fig.~\ref{fig:neggluon} we show the input PDFs at $Q_0^2=1$ GeV$^2$, and also the gluon distribution when evolved up to $Q^2=4$ GeV$^2$.  The dashed curves in Fig.~\ref{fig:neggluon} show that the inclusion of absorptive effects yield an input gluon distribution which is much less negative, whereas the input sea quark distribution is largely unaffected.
In fact, repeating the fits without the negative term in the input gluon distribution gives a description of the $F_2$ data which is almost as good ($\chisq=1.11$, compared to $1.09$ for a negative input gluon).  By contrast, without any absorptive corrections, the fit to $F_2$ is much worse without the negative term ($\chisq=1.57$, compared to $1.15$ for a negative input gluon).  We conclude that absorptive corrections remove the need for a negative gluon distribution at $Q_0^2=1$ GeV$^2$.

\begin{figure*}[!ht]
  \centering
  \includegraphics[width=0.45\textwidth]{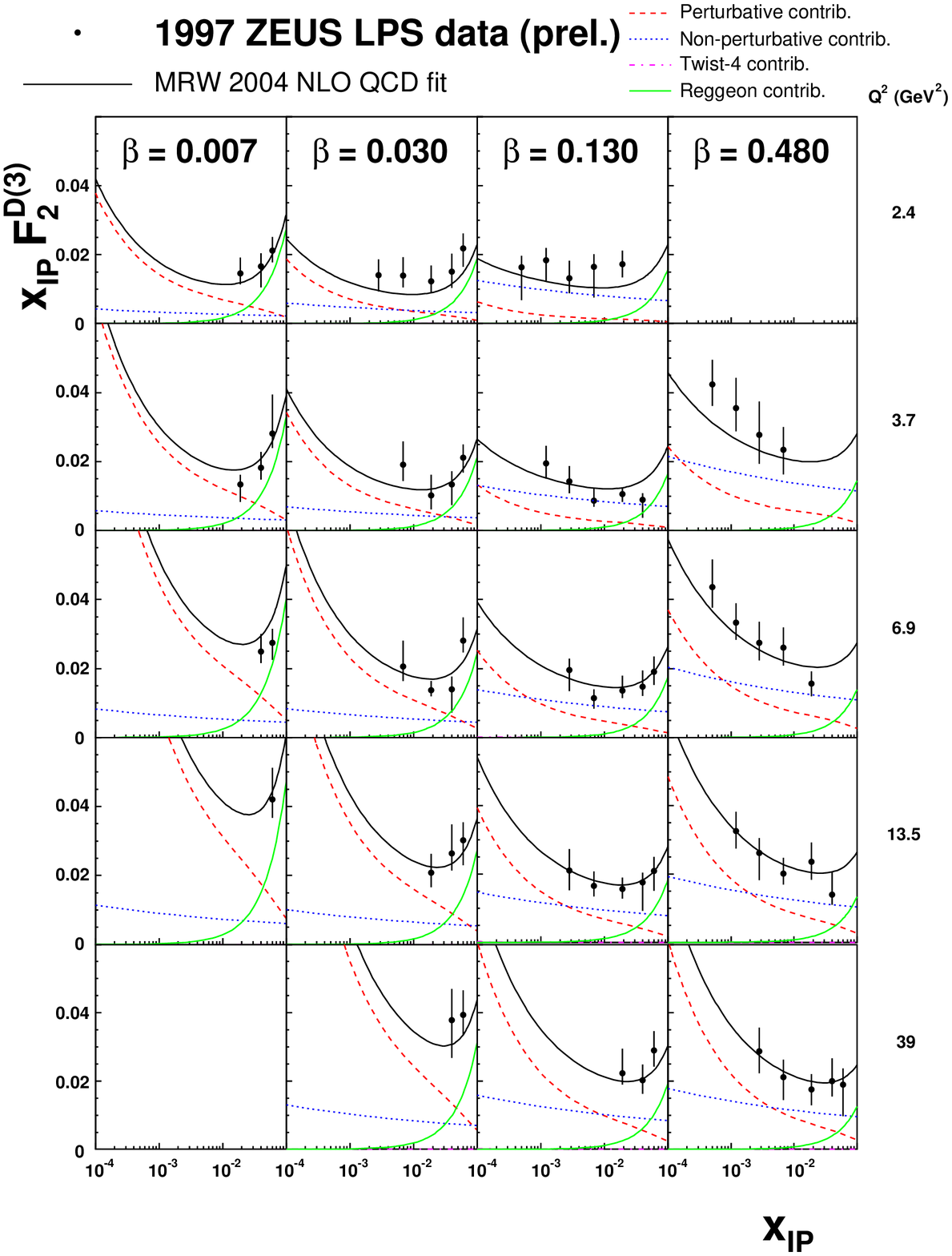}\hfill%
  \includegraphics[width=0.45\textwidth]{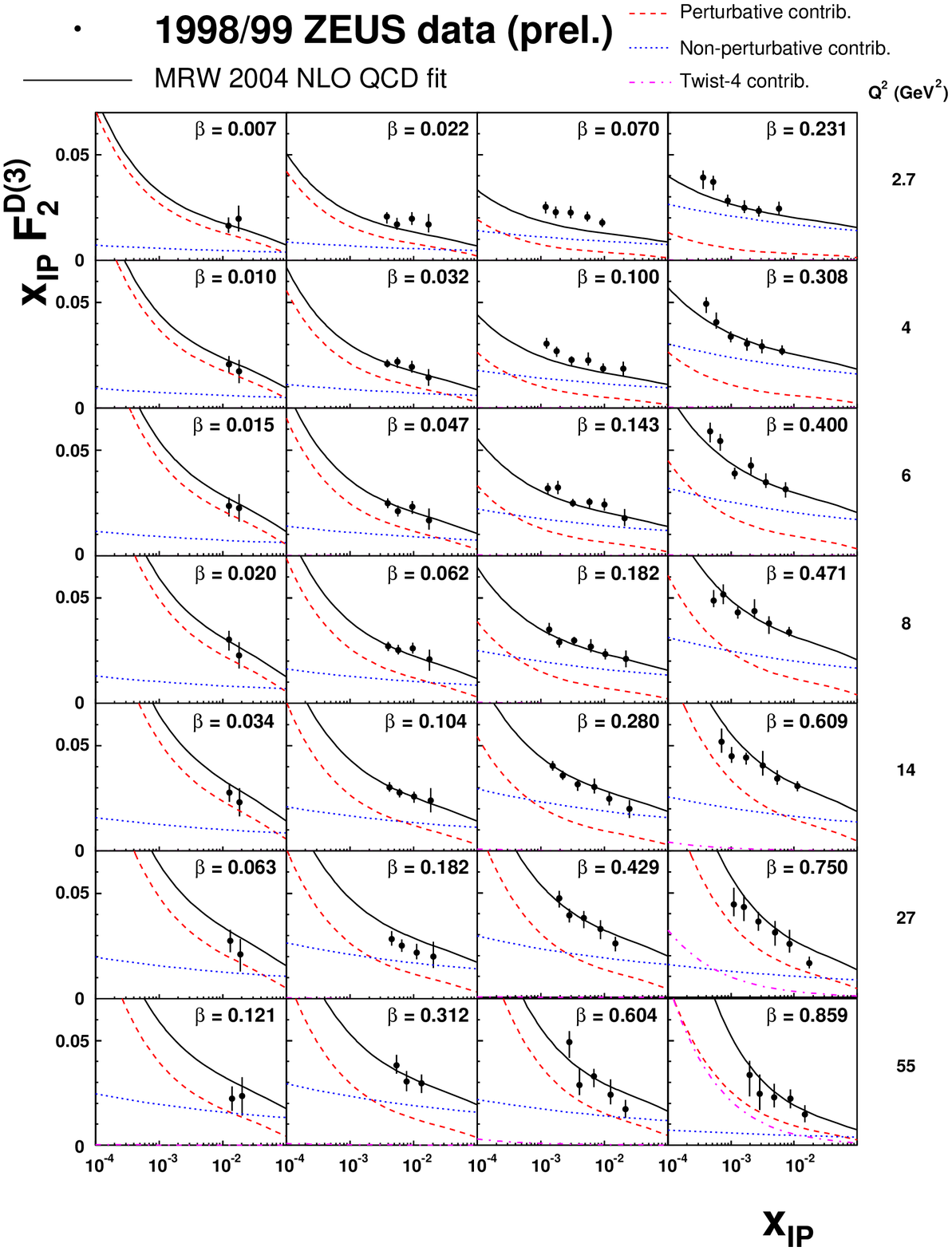}\\
  \includegraphics[width=0.82\textwidth]{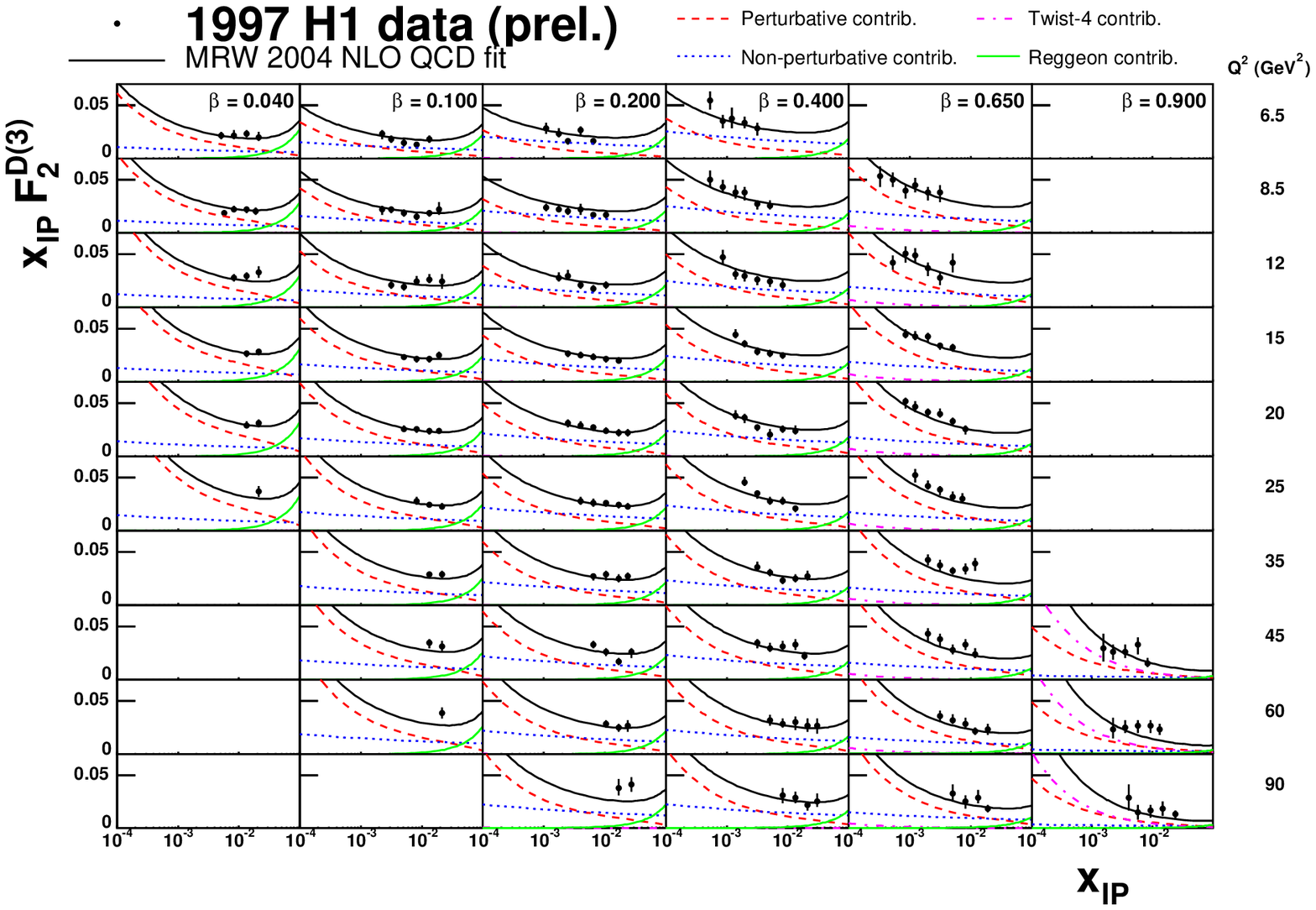}
  \caption{Description of combined preliminary ZEUS \cite{ZEUSdata} and H1 \cite{H1data} DDIS data.}
  \label{fig:f2d3data}
\end{figure*}

The corresponding description of the DDIS data, with a $\chisq=1.14$, is shown in Fig.~\ref{fig:f2d3data}.

\section{DIFFRACTIVE PDFS}
\begin{figure}[!tb]
  \centering
  \includegraphics[width=0.5\textwidth,clip]{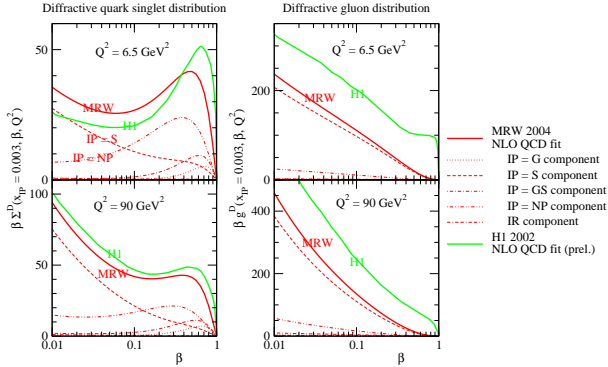}
  \caption{Diffractive PDFs, including a breakdown of the different components, compared to those obtained by H1 \cite{Schaetzel,H1data}.}
  \label{fig:dpdfs}
\end{figure}
We can extract diffractive PDFs (DPDFs) from the leading-twist contributions to the $F_2^{D(3)}$ fit shown in Fig.~\ref{fig:f2d3data}.  The diffractive quark singlet and gluon distributions are plotted in Fig.~\ref{fig:dpdfs}, and compared to those obtained by H1 \cite{Schaetzel,H1data}.  Note that the diffractive quark singlet distribution obtained by H1 has a slightly steeper $Q^2$ dependence, and hence H1 obtain a larger diffractive gluon distribution.  In addition, the smaller value of $\alpha_S(M_Z^2)$ used by H1 also enlarges their gluon density.  It has been demonstrated by H1 that their DPDFs \cite{Schaetzel,H1data} can be used to describe final state observables in DDIS, namely dijet and $D^*$ meson production cross sections \cite{Vinokurova}.  Before our DPDFs can be taken seriously we need to demonstrate this also.

\section{POMERON-LIKE SEA QUARKS BUT VALENCE-LIKE GLUONS?}

The inclusion of absorptive corrections has not removed a long-standing puzzle concerning the behaviour of PDFs at small $x$ and low scales.  We still have a valence-like gluon distribution, whereas the sea quark distribution increases with decreasing $x$.  That is, since the HERA $F_2$ data have become available, we have had a `Pomeron-like' sea quark distribution.  According to Regge theory, the high energy (small $x$) behaviour of both gluons and sea quarks is controlled by the same rightmost singularity in the complex angular momentum plane, and so we would expect $\lambda_g = \lambda_S$, where the input $xg\sim x^{-\lambda_g}$, $xS\sim x^{-\lambda_S}$.  We have studied several possibilities of obtaining a satisfactory fit with this equality imposed.  The only modification which appears consistent with the data is the inclusion of power-like corrections, which may slow down the DGLAP evolution at low $Q^2$ below about 1 or 2 ${\rm GeV}^2$.

\section{CONCLUSIONS}

In summary, we have achieved a good simultaneous description of all the DDIS and small-$\xB$ inclusive DIS data, in which the absorptive corrections in the description of the latter data have been identified and incorporated.  In this way a more physical input gluon distribution at $Q^2=1$ GeV$^2$ has been obtained, which no longer needs to be negative at small $x$.  However, there remains an outstanding dilemma in small-$x$ DIS.  Either, contrary to expectations, the non-perturbative Pomeron does not couple to gluons, or DGLAP evolution is frozen at low $Q^2$, perhaps by power corrections.  Note, however, that in both scenarios we still have the puzzle that the secondary Reggeon couples more to gluons than to sea quarks.

\end{document}